\documentclass[twocolumn]{jpsj2}

\title{%
Drude Weight of the Two-Dimensional Hubbard Model \\
\--- Reexamination of Finite-Size Effect 
in Exact Diagonalization Study \--- 
}
\catcode`\@=11
\def\simle{\mathrel{\mathpalette\@versim<}}   
\def\simge{\mathrel{\mathpalette\@versim>}}   
\def\@versim#1#2{\lower2.5pt\vbox{\baselineskip0pt \lineskip-.5pt
   \ialign{$\m@th#1\hfil##\hfil$\crcr#2\crcr\sim\crcr}}}
\catcode`\@=12

\author{%
Hiroki Nakano\thanks{E-mail address: hnakano@sci.u-hyogo.ac.jp}, Yoshinori Takahashi and Masatoshi Imada${}^{1}$
}

\inst{%
Graduate School of Material Science, University of Hyogo, 
Kouto 3-2-1, Kamiori, Ako-gun, 678-1297 \\
${}^{1}$Department of Applied Physics, University of Tokyo, Hongo 7-3-1, 
Tokyo 113-8656
}

\recdate{\today}

\abst{%
The Drude weight of the Hubbard model 
on the two-dimensional square lattice is studied 
by the exact diagonalizations applied to clusters up to 20 sites. 
We carefully examine finite-size effects by consideration of 
the appropriate shapes of clusters and 
the appropriate boundary condition beyond the limitation 
of employing only the simple periodic boundary condition.  
We successfully capture the behavior of the Drude weight 
that is proportional to the squared hole doping concentration. 
Our present result gives a consistent understanding 
of the transition between the Mott insulator and 
doped metals. 
We also find,
in the frequency dependence of the optical conductivity,
that the mid-gap incoherent part
emerges more quickly than the coherent part and rather
insensitive to the doping concentration
in accordance with the scaling of the Drude weight.
}

\kword{%
Hubbard model, Lanczos method, metal-insulator transition, 
Drude weight, optical conductivity
}

\begin{document}
\maketitle

\section{Introduction}

Metal-insulator transition is a fundamental issue 
in the condensed-matter physics. 
Particular interest has been focused 
on properties of metals near the transition between 
the Mott insulator and the metallic phase. 
Many studies have already been made from both theoretical and 
experimental points of view\cite{IFT_review}.

One of the quantities by which one can distinguish between 
metallic or insulating phases is the Drude weight $D$, 
namely the coherent component 
of the optical conductivity $\sigma (\omega)$. 
The system is metallic or insulating 
depending on $D\ne 0$ or $D=0$, respectively. 
In the Mott insulator, the Drude weight 
therefore always vanishes. 
If one dopes the Mott insulator with carriers, 
it becomes metallic 
and the Drude weight increases with further doping.

The Hubbard model is by far the simplest microscopic model 
which can show the Mott-insulator-metal transition. 
It is the most appropriate model 
for intensive theoretical studies. 
Precise determination of its properties 
by some methods with no approximations contributes much 
to the fundamental understanding 
of the metal-insulator transition. 
In spite of the simplicity, 
various aspects of typical many-body problems 
are involved in this model  
due to the presence of the Coulomb interaction. 
Many of them are still controversial even now. 
Among them is the hole-doping dependence of the Drude weight 
near the transition between the Mott insulator and the metal. 
Primary interest of this paper is the dependence 
of the two-dimensional Hubbard model 
for small doping concentration; 
this is because the Mott insulator of this model 
at half filling is considered to become metallic 
in a nontrivial way when the doping is switched on 
and because this model has been extensively studied 
as a model of high-$T_{\rm c}$ cuprates. 

Generally, it is not so easy to calculate 
such a dynamical quantity as the Drude weight 
of the strongly correlated system 
without approximations. 
One possible way is to apply the exact diagonalization method 
to finite-size clusters. 
The first work along this line 
concerning with the two-dimensional Hubbard model 
was reported by Dagotto {\it et al}\cite{Dagotto_etal}.  
Based on the treatment on the clusters up to 16 sites, 
they successfully captured a rough feature
of the Drude weight as a function 
of the hole doping concentration $\delta$. 
They tentatively concluded that $D\propto \delta$, 
although the Drude weight sometimes becomes 
negative with large absolute values 
especially near the metal-insulator transition point. 
Such negative Drude weights occur as a consequence 
of the finite-size effect. 
In order to get rid of such a finite-size effect, 
Nakano and Imada examined the properties 
under various boundary conditions in the system 
of the same size with 16 sites\cite{Nakano_Imada_1999}.  
They consequently found the concave behavior of the Drude weight 
as a function of $\delta$ in the small-$\delta$ region, 
which implies the dependence $D\propto \delta^{2}$. 
Recently, Tohyama and his co-workers treated 
larger clusters of 18 and 20 sites 
under the periodic boundary condition\cite{Tohyama_etal}. 
Their result of the Drude weight seems to suggest 
$D\propto \delta$ again.  
Thus, the issue of the $\delta$-dependence of $D$ 
is still controversial at present. 

Under these circumstances, we examine the finite-size effects 
of the Drude weight 
of clusters with various shapes and sizes up to 20 sites 
under various boundary conditions very carefully 
and do our best in minimizing the finite-size effects 
when we calculate the Drude weight of the finite-size clusters. 
The purpose of the study in this paper is 
to capture thereby a reliable exponent 
for the $\delta$-dependence of the Drude weight 
near the transition 
between the metal and the Mott insulator. 
Our present results suggest that $D\propto \delta^{2}$. 

This paper is organized as follows. 
In the next section, the model Hamiltonian and the method of 
numerical calculation are introduced. 
Section 3 is devoted to the examination 
of the finite-size effect in the non-interacting case. 
In Section 4, results in the interacting case are reported 
and discussed. 
Summary and conclusion are given in the final section. 

\section{Model and Method}

We consider the two-dimensional Hubbard model. 
Its Hamiltonian is given by
\begin{eqnarray}
{\cal H}&=&{\cal H}_{\rm hop}+{\cal H}_{\rm int}, 
\label{Hubbard_Hamiltonian}
\\
{\cal H}_{\rm hop}&=&-t \sum_{\langle i,j \rangle, \sigma}
c_{i,\sigma}^{\dagger} c_{j,\sigma}, 
\nonumber \\
{\cal H}_{\rm int}&=&
U \sum_{i} n_{i\uparrow} n_{i\downarrow}, 
\nonumber
\end{eqnarray}
where the creation (annihilation) of an electron at site $i$ 
with spin $\sigma$ is denoted by $c_{i,\sigma}^{\dagger}$ 
($c_{i,\sigma}$) and $n_{i,\sigma}$ is the corresponding 
number operator, 
namely  $n_{i,\sigma}=c_{i,\sigma}^{\dagger}c_{i,\sigma}$. 
In this paper, we treat only the nearest-neighbor hopping 
on the two-dimensional square lattice. 
Shapes for finite-size clusters will be illustrated later. 
Energies are measured in units of $t$; 
therefore, we set $t=1$ hereafter. 
The system consists of $N_{\rm e}$ electrons 
on $N_{\rm s}$ atomic sites. 
The hole doping concentration is a 
controllable parameter given by 
$\delta=1-n$, where the electron density is defined as 
$n=N_{\rm e}/N_{\rm s}$. 

We perform numerical diagonalization of finite-size clusters 
of the model (\ref{Hubbard_Hamiltonian}) 
using Lanczos algorithm. 
This method is non-biased against the effects 
of Coulomb interaction. 
The method is also useful to obtain the ground-state 
wave function with the ground-state energy $E_{\rm g}$ 
very precisely. 
On the other hand, the disadvantage of this method 
is that available system sizes are limited 
to be small. 
This is because the dimension of the Hilbert space 
grows exponentially with respect 
to the system size\cite{dimension_Hilbert_space}. 
In order to overcome this disadvantage, 
we perform parallel calculations 
in the computer systems of the distributed-memory type. 
Thereby we can treat {\it arbitrary} shapes of clusters 
up to $N_{\rm s}=20$ 
if we use appropriate supercomputer systems. 
We also employ the continued-fraction expansion 
method\cite{Cont_Frac_Method}, 
which provides us with dynamical quantities including 
the optical conductivity and the Drude weight. 
The detailed procedure will be explained after 
the examinations of cluster shapes and boundary conditions. 

\section{Finite-Size Effect for $U=0$ and Boundary Condition}

\begin{figure}[t]
\begin{center}
\includegraphics[width=8cm]{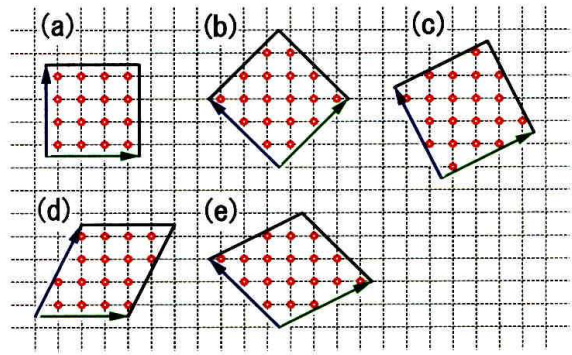}
\end{center}
\caption{Finite size clusters for $N_{\rm s}=16$, 18 and 20. 
The blue and green vectors represent 
two translational vectors for each cluster. 
The atomic sites are denoted by the red circles 
inside the quadrangle formed by 
these two vectors and the other edges denoted by black solid lines. }
\label{fig1}
\end{figure}

In this section, we examine how finite-size effects occur 
in various types of boundary conditions 
of clusters for a given $N_{\rm s}$. 
When one considers the system of two-dimensional square lattice 
by using a finite-size cluster, 
regular squares under the periodic boundary condition 
are usually adopted as in refs. \ref{Dagotto_etal} 
and \ref{Tohyama_etal}. 
Such regular squares for $N_{\rm s}=16$, 18 and 20 are 
depicted in Fig. \ref{fig1} (a), (b) and (c), respectively. 
The merit of adopting these squares is that 
the $x$ and $y$ directions are equivalent. 
When one treats such a quantity as the Drude weight, however, 
calculations based on the regular squares 
under the periodic boundary condition reveal 
a large finite-size deviation from the value 
in the thermodynamic limit. 
In ref. \ref{Dagotto_etal}, a remarkable finite-size effect 
was reported. 
In view of this situation, 
effects of other types of boundary conditions, 
namely the antiperiodic boundary condition and 
the mixed boundary condition, 
were studied in ref. \ref{Nakano_Imada_1999} 
to see whether one can reduce the finite-size effects
for the fixed $N_{\rm s}$ or not. 
Consequently, the wrong negative large Drude weights 
were successfully resolved. 
Note here that 
when one imposes the mixed boundary condition,  
the two lattice directions are no longer equivalent 
and the average of the two directions gives a reasonable result. 
This suggests that 
adhering to the equivalence of the two directions 
is not necessary. 
The two directions become inequivalent 
when one relaxes the condition 
that these quadrangles of the clusters have 
the right angle at their vertices. 
Therefore we rather 
additionally examine the parallelogram clusters 
depicted in Fig. \ref{fig1} (d) and (e) 
for $N_{\rm s}=16$ and 18 in place of squared-shaped clusters, 
respectively. 
An important point is that 
a set of possible wave number vectors
corresponding to the clusters (d) and (e) is 
different from the one corresponding 
to the clusters (a) and (b), respectively. 
The system is nevertheless still bipartite. 
For $N_{\rm s}=20$, on the other hand, 
it is impossible to preserve the bipartite nature 
when one distorts the cluster (c) so that 
it realizes a parallelogram cluster 
with a different set of possible wave number vectors. 
Then we study only one type of cluster for $N_{\rm s}=20$.
Therefore we consider these clusters (a)-(e) 
up to $N_{\rm s}=20$. 

Next, we show how to select the appropriate cluster 
under the appropriate boundary condition 
for a given $N_{\rm s}$. 
For this purpose, let us first see 
the ground-state energy $E_{\rm g}$ of the clusters (a)-(e) 
at half filling for $U=0$, presented in Table \ref{table1}. 
\begin{table}[b]
\caption{List of ground-state energy per site 
in the non-interacting case $U=0$ at half filling
for various shapes of clusters and 
various boundary conditions together 
with relative difference from the energy 
in the thermodynamic limit. 
Head BC1 (BC2) denotes the boundary condition 
along the green (blue) translational vector in Fig. \ref{fig1}. 
P and A represent the periodic boundary condition 
and the antiperiodic boundary condition, respectively. 
Note here that the boundary conditions 
giving the same energy for each cluster correspond 
to the identical set of possible wave number vectors. 
Marked cases in last column with head PW show 
the selected clusters under the selected boundary condition 
in which we actually perform calculations in the present work. 
}
\label{table1}
\begin{center}
\begin{tabular}{@{\hspace{\tabcolsep}\extracolsep{\fill}}c|cc|lc|c} 
\hline
cluster & BC1 & BC2 & $-E_{\rm g}/N_{\rm s}$ & 
$(E_{\infty}-E_{\rm g})/E_{\infty}$  & PW  
\\
\hline
(a) & P & P & 1.5     & $-$0.0747 & \\ 
    & A & P & 1.70711 & $+$0.0530 & \\ 
    & P & A & 1.70711 & $+$0.0530 & \\ 
    & A & A & 1.41421 & $-$0.1276 & \\ 
\hline
(b) & P & P & 1.77778 & $+$0.0966 & \\ 
    & A & P & 1.53960 & $-$0.0503 & \\ 
    & P & A & 1.53960 & $-$0.0503 & \\ 
    & A & A & 1.33333 & $-$0.1775 & \\ 
\hline
(c) & P & P & 1.69443 & $+$0.0452 & \\ 
    & A & P & 1.65065 & $+$0.0182 & $\bullet$ \\ 
    & P & A & 1.65065 & $+$0.0182 & \\ 
    & A & A & 1.52169 & $-$0.0613 & \\ 
\hline
(d) & P & P & 1.70711 & $+$0.0530 & \\ 
    & A & P & 1.63099 & $+$0.0061 & $\bullet$ \\ 
    & P & A & 1.5     & $-$0.0747 & \\ 
    & A & A & 1.63099 & $+$0.0061 & \\ 
\hline
(e) & P & P & 1.72417 & $+$0.0636 & \\ 
    & A & P & 1.64519 & $+$0.0148 & $\bullet$ \\ 
    & P & A & 1.51621 & $-$0.0647 & \\ 
    & A & A & 1.44675 & $-$0.1076 & \\ 
\hline
\end{tabular}
\end{center}
\end{table}
Note that this quantity is proportional to the Drude weight 
for $U=0$. 
A noticeable point 
of the results of clusters (a), (b) and (c) 
under the periodic boundary condition 
is that 
the relative difference $(E_{\infty}-E_{\rm g})/E_{\infty}$ 
does not monotonically converge as $N_{\rm s}$ is increased, 
where $E_{\infty}$ denotes the ground-state energy 
in the limit of $N_{\rm s}\rightarrow\infty$. 
This suggests that 
the system size $N_{\rm s}$ up to 20 is not free 
from the finite-size effect as far as we stick 
to the periodic boundary condition.  
We have thus to be careful in choosing 
an appropriate cluster under an appropriate boundary condition
for a given $N_{\rm s}$. 
Let us now compare the difference 
$(E_{\infty}-E_{\rm g})/E_{\infty}$ 
of all the clusters (a)-(e) with various boundary conditions. 
One can find the best cluster and boundary condition 
from among the cases shown in Table \ref{table1} 
so that the absolute value of the difference is the smallest. 
For $N_{\rm s}=16$ and 20, two boundary conditions give 
the smallest one. 
Sets of possible wave number vectors of the two cases 
for each $N_{\rm s}$ are equivalent. 
Thus, the same result will be obtained 
if one selects either of the two. 

\begin{figure}[t]
\begin{center}
\includegraphics[width=8cm]{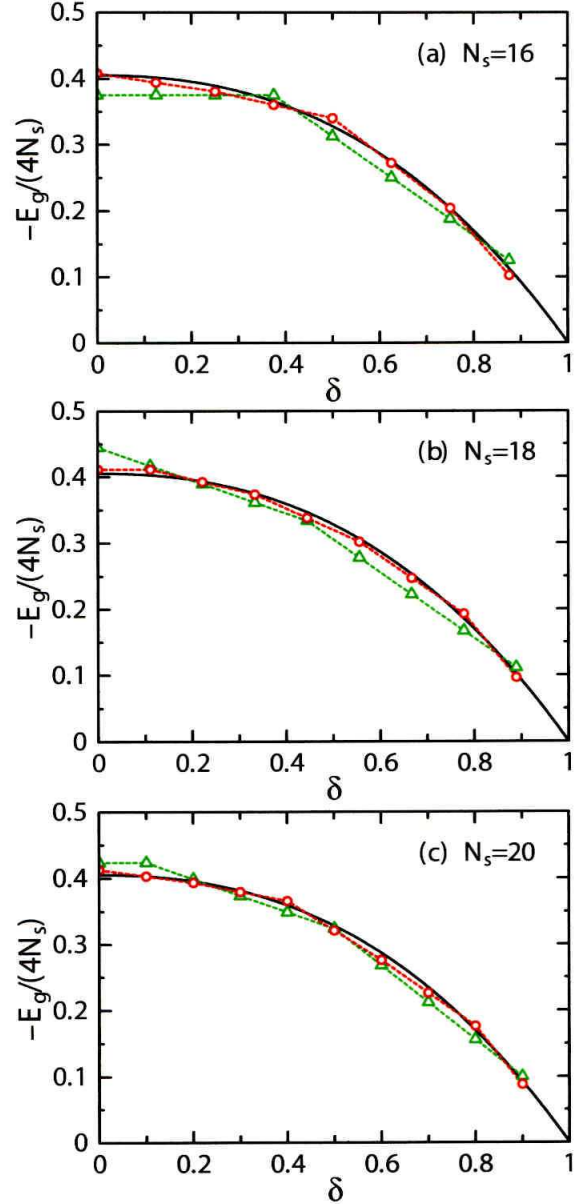}
\end{center}
\caption{Ground-state energies as function of $\delta$ 
in non-interacting case 
for given $N_{\rm s}$ presented in the form of 
$-E_{\rm g}/(4N_{\rm s})$. 
This quantity corresponds to $D$ for $U=0$. 
Green triangles represent data for the regular square 
under the periodic boundary condition. 
Red circles represent data 
for the appropriate cluster 
under the appropriate boundary condition 
based on the analysis in Table \ref{table1}. 
The solid curve denotes the corresponding quantity 
for the thermodynamic limit. 
}
\label{fig2}
\end{figure}

Let us compare the $\delta$-dependence 
of the non-interacting $E_{\rm g}$ between two cases, 
the one of the most appropriate cluster 
under the most appropriate boundary condition 
with the best $E_{\rm g}$ at half filling 
and the other of the regular square 
under the periodic boundary condition; 
results are depicted in Fig. \ref{fig2}. 
One can easily observe the results of the former case 
fall closer to the curve of the thermodynamic limit 
in the whole region of $\delta$, 
irrespective of the size $N_{\rm s}$.  
On the other hand, the latter regular-square case 
under the periodic boundary condition gives 
significant overestimates 
particularly at ($N_{\rm s}$, $N_{\rm e}$) 
= (18, 18) and (20, 18). 
This strongly suggests that 
the Drude weights of these cases will be possibly overestimated 
even when $U$ is considerably large. 
It also suggests that 
if the intrinsic Drude weight shows the concave behavior 
as a function of $\delta$
for large $U$, 
this overestimation will possibly cancel this behavior. 
Figure \ref{fig2} also provides 
another important finite-size effect 
that finite-size data show nonzero curvature 
in their $\delta$-dependence only at $\delta$ corresponding 
to the closed electronic shell structure. 
At $\delta$ for the open electronic shell structure, 
on the other hand, $E_{\rm g}$ follows linear dependence. 
The linear dependence from this origin is likely to
survive more or less even for $U>0$. 
This is, in fact, confirmed in the quantum Monte Carlo study 
at nonzero $U$~\cite{Furukawa_Imada_QMC_1992}.
Thus, it is difficult 
to judge whether the linear $\delta$-dependence 
revealed from numerical studies 
in the interacting case in the region of the open shell structure 
is intrinsic or is simply attributed
to this particular electronic structure. 
In view of this situation, 
it is desirable to confine our analysis to 
the data 
at $\delta$ for the closed shell structure 
in order to extract the intrinsic behavior.  
In this paper, therefore, 
we take the appropriate clusters under the 
the appropriate boundary condition for each $N_{\rm s}$ 
(shown in Table \ref{table1}) and adopt 
data for the filling forming the closed shell structure. 

In the selected clusters and boundary conditions, 
the equivalence of the two directions denoted 
by $\alpha$ ($=x,y$) does not hold, 
as we have already mentioned in the above. 
Then, we take the direction average according to 
the procedure used in ref. \ref{Nakano_Imada_1999}. 
The optical conductivity is given by 
$\sigma (\omega)= 
2\pi  D \delta(\omega) 
+\sigma^{\rm reg}(\omega)$.  
Here, the second term is the regular part of the 
optical conductivity, which is also called the incoherent part. 
The regular part $\sigma^{\rm reg}(\omega)$ is obtained 
by the average 
$\sigma^{\rm reg}(\omega)=
[\sigma^{\rm reg}_{x}(\omega)+\sigma^{\rm reg}_{y}(\omega)]/2$; 
each of $\sigma^{\rm reg}_{\alpha}(\omega)$ given by 
\begin{equation}
\sigma^{\rm reg}_{\alpha}(\omega)= 
\frac{\pi }{N_{\rm s}}  
\sum_{\ell(\ne 0)} 
\frac{|\langle \ell |j_{\alpha}|0\rangle |^2}{E_{\ell}-E_{0}} 
\delta(\omega-E_{\ell}+E_{0}), 
\label{Kubo_formula}
\end{equation}
is calculated by the continued-fraction expansion method. 
Here, $j_{\alpha}$ is a current operator 
along the $\alpha$-direction defined as 
\begin{equation}
j_{\alpha}=-{\rm i} 
\sum_{i,\sigma} t (
 c_{i,\sigma}^{\dagger} c_{i+\delta_{\alpha},\sigma} 
-c_{i+\delta_{\alpha},\sigma}^{\dagger} c_{i,\sigma} ),
\end{equation}
where $\delta_{\alpha}$ is the unit vector 
along the $\alpha$-direction, 
$|\ell\rangle$ represents an eigenstate 
with the energy eigenvalue $E_{\ell}$.  
Note that the ground state is described by $\ell=0$. 
One can obtain the Drude weight $D$ 
from the combination of $\sigma(\omega)$ and the sum rule 
\begin{equation}
\int_{0}^{\infty}\sigma(\omega){\rm d}\omega = \pi K, 
\end{equation}
where $-4K$ represents the kinetic energy per site,
namely $-4K=\langle {\cal H}_{\rm hop} \rangle/N_{\rm s}$. 
For $U=0$, the incoherent part is absent; 
therefore one has $D=K$. 
It is widely known that for $U>0$, 
the incoherent part appears. 
When $U$ is large enough, one can easily recognize 
in $\sigma(\omega)$ the responses toward the upper-Hubbard band. 
In such a case, one can definitely determine  
a frequency just below the upper-Hubbard band to be 
$\omega_{\rm c}$; 
we also calculate definite integral 
\begin{equation}
N_{\rm eff} = \frac{1}{\pi} \int_{0}^{\omega_{\rm c}} 
\sigma(\omega) {\rm d} \omega , 
\end{equation}
which is called an effective carrier density. 

\begin{figure}[t]
\begin{center}
\includegraphics[width=8cm]{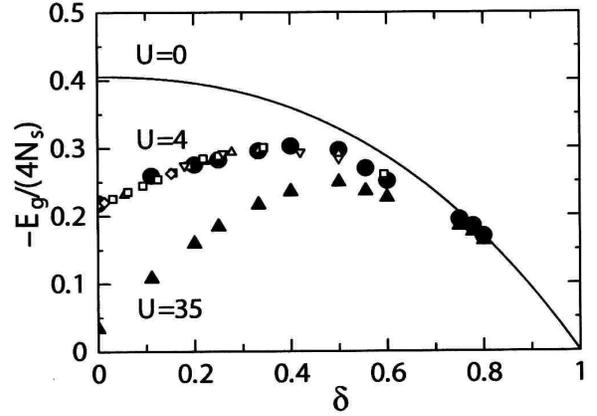}
\end{center}
\caption{Ground-state energies as function of $\delta$ 
in interacting cases 
presented in the form of $-E_{\rm g}/(4N_{\rm s})$. 
The solid curve denotes $-E_{\rm g}/(4N_{\rm s})$ for $U=0$ 
for the thermodynamic limit. 
Closed circles and closed triangles denote the present results 
from exact diagonalization calculations for $U=4$ and $U=35$, 
respectively.  
Open symbols (triangles, squares, reversed triangles and 
diamonds) denote results of $N_{\rm s}=6\times 6$, $8\times 8$, 
$10\times 10$ and $12\times 12$ for $U=4$
from the quantum Monte Carlo simulations, 
respectively\cite{Furukawa_Imada_QMC_1992}. 
}
\label{fig3}
\end{figure}

As for the validity of extending the present procedure 
to cases with finite $U$, 
there is no convincing theoretical argument on the condition 
of reducing the finite-size effect. 
Though not reliable in a strict and rigorous sense, 
it is at least better to start from situations 
where the effects are reduced 
in the case of $U=0$ from the continuity of behavior 
as the function of $U$. 
Another evidence for the validity of the present procedure 
is given from our examination of $E_{\rm g}$ for $U>0$; 
the results are depicted in Fig. \ref{fig3}. 
Although all the possible data of $N_{\rm s}=16$, 18 and 20 
are shown there with the same closed symbol for each case 
of $U>0$, $\delta$-dependence of $E_{\rm g}$ is fairly smooth.
This fact implies that finite-size deviations 
of the present data are small. 
One can actually confirm that the present results 
for $U=4$ agree well with the results 
from the quantum Monte Carlo 
simulations\cite{Furukawa_Imada_QMC_1992} on larger systems 
up to $N_{\rm s}=12\times12$. 
Therefore, it is reasonable to presume that 
the finite-size effects in $D$ and $N_{\rm eff}$ 
based on the present procedure are also small 
if the smooth $\delta$-dependence of $D$ and $N_{\rm eff}$ 
is confirmed irrespective of the size $N_{\rm s}$. 

In this paper, we take three cases of $U=10$, 20 and 35. 
In the system with a nonzero $U$ weaker than $U=10$, 
it is difficult to distinguish 
the upper-Hubbard band and the lower-Hubbard band 
in the responses of $\sigma(\omega)$; 
the estimation of $N_{\rm eff}$ is not available. 
On the other hand, in the system for $U>35$, 
the ground state is possibly ferromagnetic near half filling. 
Once the ferromagnetism appears, 
it is too difficult to study 
the scaling of $D$ near half filling.  

\section{Results for Interacting Case and Discussions}

\begin{figure}[t]
\begin{center}
\includegraphics[width=8cm]{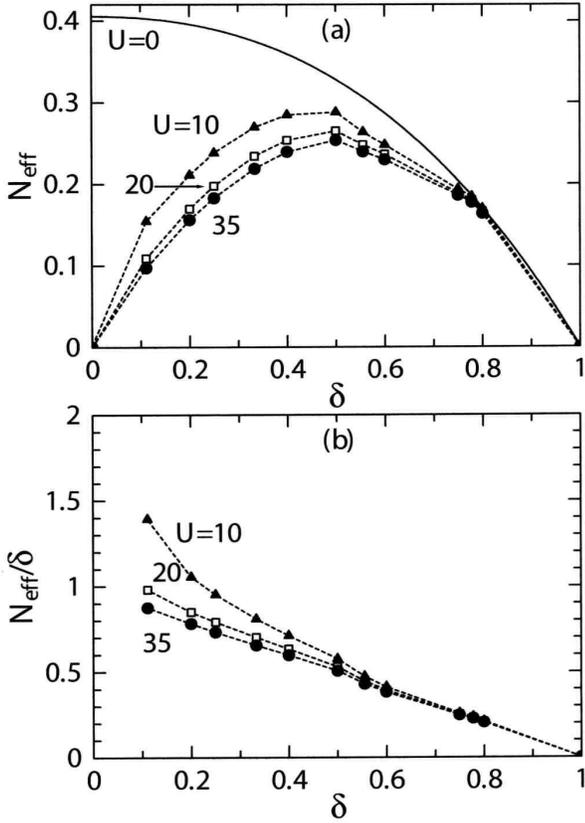}
\end{center}
\caption{$\delta$-dependence of $N_{\rm eff}$ in (a) and 
$N_{\rm eff}/\delta$ in (b).  
Closed triangles, open squares and closed circles denote data 
for $U=10$, 20 and 35, respectively. 
In (a), total weight for $U=0$ is also depicted 
by the solid curve for comparison. 
}
\label{fig4}
\end{figure}

We show numerical results for the $\delta$-dependence 
of $N_{\rm eff}$ connected by lines 
for interacting cases in Fig. \ref{fig4}. 
The $\delta$-dependence of the Drude weight is later shown. 
Although the data consist of those of various system sizes 
for each case of $U$, $\delta$-dependence of them is fairly 
smooth, independent of the value $U$, as shown 
in Fig. \ref{fig4}(a). 
It clearly indicates that 
deviations of our numerical results 
due to the finite-size effect are very small 
for every $U$ that we treat. 
One can also observe that 
as $\delta$ is decreased 
toward the half-filling case $\delta=0$, 
$N_{\rm eff}$ goes to a very small value 
for each of non-zero $U$. 
The values of $N_{\rm eff}$ at half filling 
are very close to the corresponding $D$. 
This is because no significant responses 
within $0<\omega<\omega_{\rm c}$ are observed 
in the optical conductivity; 
this behavior has already been reported 
in all the previous 
works\cite{Dagotto_etal,Tohyama_etal,Nakano_Imada_1999} and 
will be confirmed later in the present study. 
Then, we will later come back to these values 
of $N_{\rm eff}$ at half filling 
in the discussion of the Drude weight. 
In order to examine the $\delta$-dependence of $N_{\rm eff}$ 
near half filling, 
Fig. \ref{fig4}(b) depicts the $\delta$-dependence 
of $N_{\rm eff}/\delta$ instead of raw $N_{\rm eff}$. 
One can observe the linear behavior irrespective 
of the value of $U$ 
in the wide region of $\delta$. 
They extrapolate to finite intercepts of the 
$N_{\rm eff}/\delta$ axis, 
resulting in the following dependence  
\begin{equation}
N_{\rm eff} \propto \delta ,
\label{dependence_N_eff}
\end{equation}
near half filling $\delta=0$. 
Note that the above dependence (\ref{dependence_N_eff}) is in
agreement with all the previous 
studies\cite{Dagotto_etal,Tohyama_etal,Nakano_Imada_1999}.

\begin{figure}[t]
\begin{center}
\includegraphics[width=8cm]{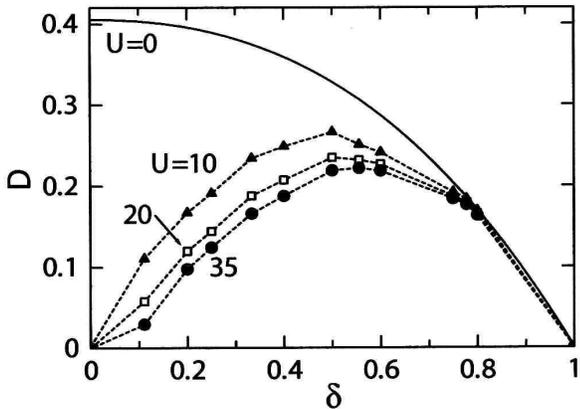}
\end{center}
\caption{$\delta$-dependence of Drude weight 
for $U=10$, 20 and 35. 
Notations for closed triangles, open squares and closed circles are 
the same as those in Fig. \ref{fig4}. 
For comparison, solid curve represents $D$ for $U=0$, 
which is the same as the total weight. 
}
\label{fig5}
\end{figure}
We depict $\delta$-dependence of the Drude weight 
in Fig. \ref{fig5}. 
Note that the Drude weight at $\delta=0$ 
vanishes for $U>0$ in the thermodynamic limit. 
All the values for various $U$ are so small, 
less than 0.003, that we cannot distinguish their 
symbols from the origin as shown 
in Fig. \ref{fig4}(a) and \ref{fig5}, 
where our data of the half-filling case 
for $N_{\rm s}=20$ are plotted. 
For $N_{\rm s}=16$,
we obtain $D\sim 0.009$ for $U=10$ at most;
since its difference from the data of $N_{\rm s}=20$ is
only within the symbol size, it is not shown in the figure.
On the other hand,
ref. \ref{Tohyama_etal} reported the Drude weight of about 0.02,
which is substantially larger than ours,
for the cluster (c) with $N_{\rm s}=20$
under the periodic boundary condition.
Our present result is closer to the correct value of zero  
in spite of the smaller size $N_{\rm s}$, 
which indicates superiority of our present calculations 
in reproducing well the insulating state at half filling. 

In the metallic region for $\delta > 0$, 
all the data obtained from studies on various system sizes are 
presented in Figs. \ref{fig4}(a) and \ref{fig5}
according to the discussion in section 3. 
Recall here that refs. \ref{Dagotto_etal} and 
\ref{Tohyama_etal} reported anomalous and 
discontinuous deviations of $D$ at some particular hole 
concentrations, resulting 
from a finite-size effect\cite{comment_negative_D}. 
On the other hand, 
the  present data for all the values of $\delta$ 
in Fig. \ref{fig5} show smooth variation on the whole, 
though they consist of mixture for various $N_{\rm s}$. 
It indicates that the finite-size effect 
is well eliminated in our estimates. 
As for the $\delta$-dependence of the Drude weight, 
the convex behavior is apparent 
in all the region of $\delta$ for $U=10$. 
For $U=20$ and 35, on the other hand, the concave behavior 
occurs near half filling. 
The change in this $\delta$-dependence with increasing $U$ 
will be discussed below. 

\begin{figure}[t]
\begin{center}
\includegraphics[width=8cm]{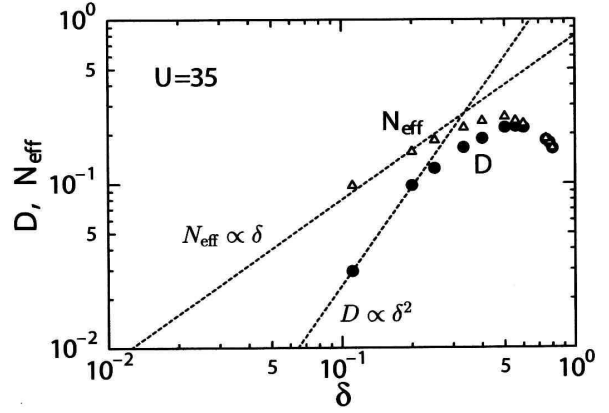}
\end{center}
\caption{Log-log plot of $D$ and $N_{\rm eff}$ 
as function of $\delta$ for $U=35$. 
Circles and triangles denotes $D$ and $N_{\rm eff}$, 
respectively. 
}
\label{fig6}
\end{figure}
In order to confirm the concave behavior of $D$, 
we plot the data of $D$ on the logarithmic scale 
in Fig. \ref{fig6} together with $N_{\rm eff}$  for $U=35$. 
Our data of $D$ near half filling indicate the dependence 
\begin{equation}
D \propto \delta^{2}. 
\label{dependence_incoherent_D}
\end{equation}
This dependence is in contrast to the linear dependence 
of $N_{\rm eff}$ of eq. (\ref{dependence_N_eff}). 
\begin{figure}[t]
\begin{center}
\includegraphics[width=8cm]{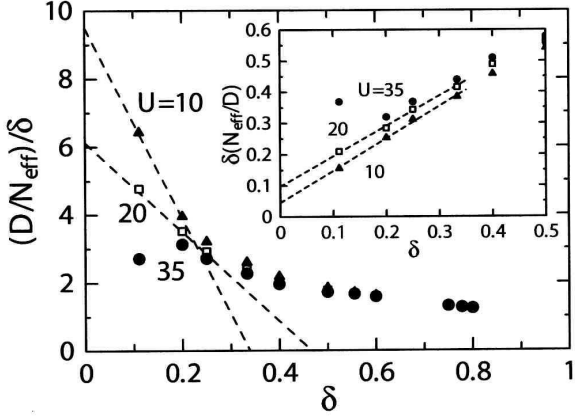}
\end{center}
\caption{$\delta$-dependence of $D/(N_{\rm eff}\delta)$ and 
its inverse. 
Notations for closed triangles, open squares and closed circles are 
the same as those in Fig. \ref{fig4}. 
}
\label{fig7}
\end{figure}
For more detailed analysis of the dependence of $D$ 
in the region of small $\delta$, 
let us next examine the $\delta$-dependence 
of the ratio $D/N_{\rm eff}$. 
The corresponding quantity was also studied 
in the two-dimensional $t$-$J$ model\cite{Tsunetsugu_Imada_tJ}. 
The present results are shown in Fig. \ref{fig7} 
in the form of $(D/N_{\rm eff})/\delta$. 
For $U=10$ and 20, as $\delta$ is decreased 
toward half filling, 
the quantity $(D/N_{\rm eff})/\delta$ shows steep increase. 
For $U=35$, on the other hand, no such behavior is observed. 
The increasing behavior for $U=10$ and 20 is examined in the plot 
of the inverse of $(D/N_{\rm eff})/\delta$ 
shown in the inset of Fig.~\ref{fig7}. 
Our data for $U=10$ and 20 reveal linear dependence 
in a wide region of $\delta$. 
For $U=35$, a similar linear dependence is observed 
in $0.2 \simle \delta \simle 0.5$ together 
with a deviation from it in the smaller-$\delta$ region. 
Each of the linear extrapolation of plots 
intersects at a non-zero intercept with the axis of ordinates 
in the limit of $\delta\rightarrow 0$. 
If one takes its weak $\delta$-dependence for $U=35$ 
into account, it is natural to consider that 
$[(D/N_{\rm eff})/\delta]^{-1}$ for various $U$ stays finite 
in the limit of $\delta\rightarrow 0$ as common behavior; 
there is no sign of vanishing $[(D/N_{\rm eff})/\delta]^{-1}$ 
from our data of finite size clusters. 
The above behavior, together with eq. (\ref{dependence_N_eff}), 
thus leads us to the scaling (\ref{dependence_incoherent_D}) 
valid not only for $U=35$ but also for $U=10$ and 20. 
This analysis is particularly important since it suggests 
the validity of the scaling (\ref{dependence_incoherent_D}) 
even for $U=10$ in spite of the fact that 
the $\delta$-dependence of $D$ for this $U$ in Fig. \ref{fig5} 
does not show concave behavior which appears for $U=20$ and 35. 

Let us now discuss the present result from the viewpoint of 
the scaling theory developed by Imada, a general theory 
of the metal-insulator transition\cite{Imada_scaling_1995}. 
This helps us to justify 
the dependence (\ref{dependence_incoherent_D}) in relation 
to other behavior of the two-dimensional Hubbard model. 
According to the scaling theory, the Drude weight 
shows the following critical behavior 
near the metal-insulator transition, 
\begin{equation}
D \propto \delta^{1+(z-2)/d},
\label{critical_behavior_D}
\end{equation}
where $z$ denotes the dynamical exponent and 
$d$ is the spatial dimension. 
Since $d=2$ in the present problem, 
one obtains $z=4$ from (\ref{dependence_incoherent_D}). 
The scaling theory also predicts the critical behavior of 
the compressibility near the transition,
\begin{equation}
\kappa \propto \delta^{1-z/d}. 
\end{equation}
Substitution of $z=4$ and $d=2$ in the above dependence 
of $\kappa$ leads to $\kappa\propto \delta^{-1}$. 
The diverging behavior of $\kappa$ 
was reported by the quantum Monte Carlo (QMC) 
simulations\cite{Furukawa_Imada_QMC_1992,Furukawa_Imada_QMC_1993} 
performed on the same model with larger system sizes 
not yet treated by exact-diagonalization studies. 
The QMC calculation\cite{Assaad_Imada_QMC_1996} 
also demonstrated that the localization length 
suggests the divergence $\xi_{\rm l}\propto 
|\mu-\mu_{\rm c}|^{-\nu}$ with $\nu=1/4$ 
as the chemical potential approaches 
the charge gap $\mu_{\rm c}$ from the insulating side. 
The scaling theory indicates $z\nu=1$, 
which also leads to the exponent $z=4$. 
Therefore, our present result is consistent 
with these QMC results in views of the scaling theory. 
Recent study of the metal-insulator 
transition\cite{Imada_scaling_2005a,Imada_scaling_2005b,
Misawa_2006} 
clarified that the exponent $z=4$ appears 
at the marginal quantum critical point 
between the continuous metal-insulator transition 
with $T_{\rm c}=0$ and 
the discontinuous transition with nonzero $T_{\rm c}$. 
The appearance of the same marginal exponent $z=4$ 
was also found in the one-dimensional Hubbard model 
with next-nearest-neighbor hopping\cite{Nakano_Takahashi_MI_2004,
Nakano_Takahashi_Imada_CS_2005}. 
The unified understanding of the marginal point occurring 
in the one- and two-dimensional cases is the subject 
for future studies. 
Although it is difficult
to determine the marginal quantum critical point precisely,
the present results suggest that the critical region
of the marginal point is wide extending
to all the values of $U$ we studied.
This allows us to capture successfully the distinct signature
characteristic to the marginal point.

\begin{figure}[t]
\begin{center}
\includegraphics[width=8cm]{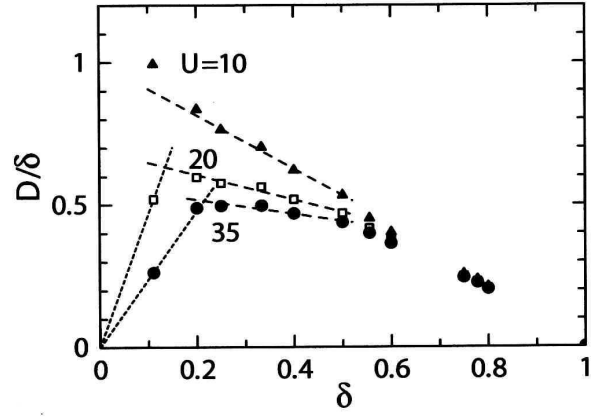}
\end{center}
\caption{$\delta$-dependence of $D/\delta$. 
Notations for closed triangles, open squares and closed circles are 
the same as those in Fig. \ref{fig4}. 
Broken lines are the fitting lines 
in $0.25 \le \delta \le 0.5$. 
Dotted lines are guides for eyes 
in the assumption of $D\propto\delta^{2}$. }
\label{fig8}
\end{figure}

Next, let us consider the width of the critical region 
of $\delta$ in which the behavior (\ref{dependence_incoherent_D}) 
is observed. 
For this purpose, we depict 
the $\delta$-dependence of $D/\delta$ in Fig. \ref{fig8}. 
In $0.25 \simle \delta \simle 0.5$,
one can observe $U$-irrelevant linear behavior. 
In Fig. \ref{fig8}, 
we fit our numerical data in $0.25 \le \delta \le 0.5$ 
with broken lines. 
For $U=20$ and 35, our data near half filling 
digress from the broken lines. 
In particular, for $U=35$, two points shows clear deviation 
toward the origin in Fig. \ref{fig8}, 
while one point for $U=20$. 
Here, let us draw the dotted guide lines 
for these digressing data 
by assuming the present result (\ref{dependence_incoherent_D}). 
The dotted and the broken lines cross with each other 
for each $U$. 
One can regard the crossing point for each $U$ 
as a rough estimation of the boundary 
of the critical region. 
Then, the estimated boundaries are 
$\delta \sim 0.14$ and 0.22 
for $U=20$ and 35, respectively. 
Though crude are these estimations, 
this is the first study to report the boundary. 
Calculations of systems with larger $N_{\rm s}$ 
with proper treatment of finite-size effects 
will make the precise estimations possible 
in future studies. 
For $U=10$, there are no data points showing 
digressing towards the origin. 
If the $\delta^2$ dependence of $D$, 
i.e. eq. (\ref{dependence_incoherent_D}), 
is true, then the above fact 
suggests that the critical region for $U=10$ is narrower 
than $\delta < 1/9$. 
This is consistent with the rapid decrease 
of the estimated boundary as $U$ decreases from 35 to 20. 
The doping concentration at the extrapolated boundary 
for $U=10$ may be very small.

\begin{figure}[t]
\begin{center}
\includegraphics[width=8cm]{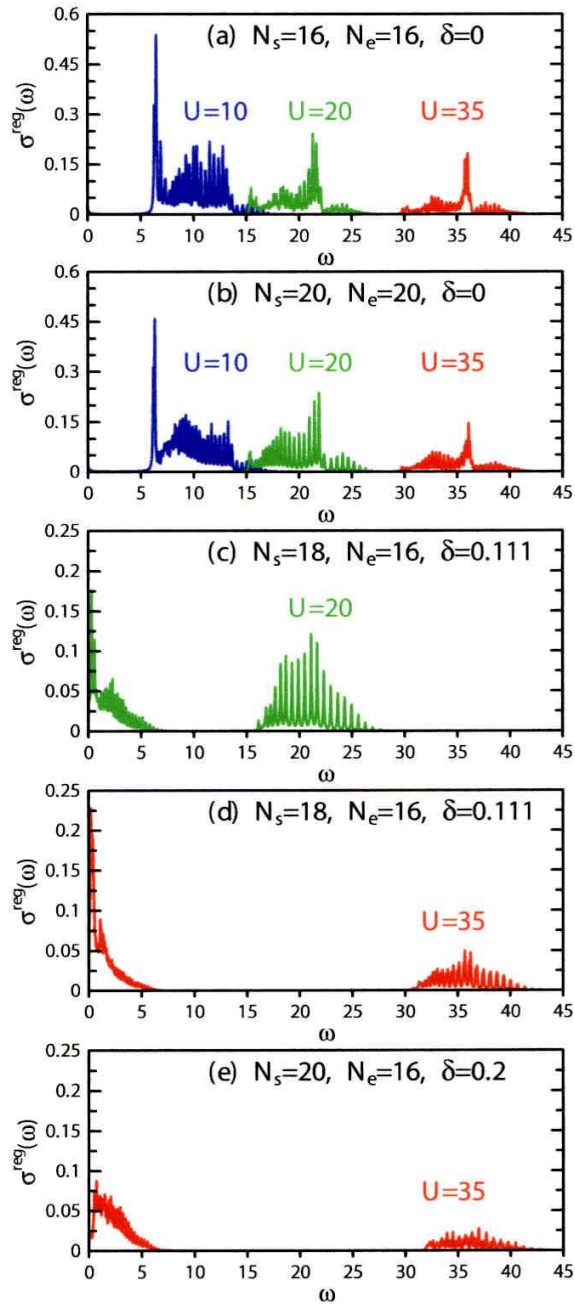}
\end{center}
\caption{Frequency dependence of regular part 
of the optical conductivity 
in (a) for $N_{\rm s}=16$ and $N_{\rm e}=16$, namely $\delta=0$, 
(b) for $N_{\rm s}=20$ and $N_{\rm e}=20$, namely $\delta=0$, 
(c) for $N_{\rm s}=18$ and $N_{\rm e}=16$, namely $\delta=1/9$, 
(d) for $N_{\rm s}=18$ and $N_{\rm e}=16$, namely $\delta=1/9$ 
and (e) for $N_{\rm s}=20$ and $N_{\rm e}=16$, 
namely $\delta=0.2$. 
Blue, green and red lines denote the cases for 
$U=10$, $U=20$ and $U=35$, respectively. 
Delta functions are broadened with width of $\eta=0.05$. 
}
\label{fig9}
\end{figure}

Let us discuss the frequency dependence 
of $\sigma^{\rm reg}(\omega)$ obtained by eq. (\ref{Kubo_formula})
for the insulating and metallic cases where the system 
is in the critical region; 
we present our results in Fig. \ref{fig9}. 
Figures \ref{fig9}(a) and \ref{fig9}(b) depict the results 
for insulating cases. 
The structures 
of $\sigma^{\rm reg}(\omega)$ between $N_{\rm s}=16$ and 
$N_{\rm s}=20$ are similar for each $U$. 
This fact suggests that finite-size effects are 
rather irrelevant for the structure 
of $\sigma^{\rm reg}(\omega)$. 
For each $U$, broad responses corresponding to the 
transitions to the upper-Hubbard band appear around $\omega=U$. 
For $U=10$, a sharp peak emerges at the lower edge of the band. 
According to ref. \ref{Tohyama_etal}, 
this sharp peak is ascribed to the spin-polaron formation 
in the photoexcited state, 
and the weight of the peak becomes smaller 
for larger $N_{\rm s}$. 
On the other hand, our sharp peak for $U=10$ 
does not show any significant $N_{\rm s}$ dependence, 
which implies that it will survive in the thermodynamic limit, 
although the peaks get rapidly weaker for larger $U=20$ and 35. 
The broad responses are unaffected by the change in $U$ 
between $U=20$ and 35 except that the center of responses 
moves with $U$. 
Figures \ref{fig9}(c), \ref{fig9}(d) and \ref{fig9}(e) depict 
the metallic cases; 
one can observe 
not only the high-frequency spectrum due to 
the transitions to the upper-Hubbard band 
but also a pronounced structure in the low-energy region 
below the Mott gap. 
First, let us see the transitions to the upper-Hubbard band. 
By switching $\delta$ on, 
these transitions gradually lose their weights 
by transferring them to the lower-energy structure, 
as already been reported in refs. \ref{Dagotto_etal}, 
\ref{Nakano_Imada_1999} and \ref{Tohyama_etal}. 
Next, let us see the structure in the low-frequency region. 
We find strong responses growing rapidly towards $\omega=0$ 
with decreasing $\omega$. 
It is noticeable that the shape seems to be a common property 
irrespective of $U$ and $\delta$ 
within the present parameter ranges.  
The same shapes are produced in the present results 
of both $N_{\rm s}=18$ and 20. 
The $\omega$-dependence of the shape reminds us of 
the $1/\omega$ tail and 
the mid-IR peak observed in the experiments 
of high-$T_{\rm c}$ compound\cite{LaSrCuO}. 

\begin{figure}[t]
\begin{center}
\includegraphics[width=8cm]{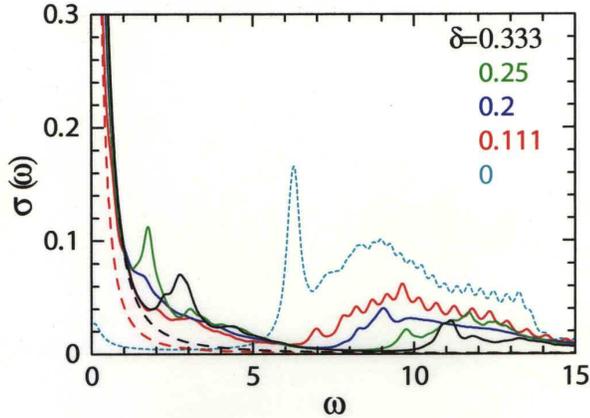}
\end{center}
\caption{Frequency dependence of optical conductivities 
for $U=10$. 
The insulating case of 
$N_{\rm s}=20$ and $N_{\rm e}=20$ is denoted 
by the sky-blue dotted line.
Red, deep-blue, green and black solid lines correspond 
to the metallic cases of 
$N_{\rm s}=18$ and $N_{\rm e}=16$, 
$N_{\rm s}=20$ and $N_{\rm e}=16$, 
$N_{\rm s}=16$ and $N_{\rm e}=12$ and 
$N_{\rm s}=18$ and $N_{\rm e}=12$, respectively.  
Delta functions are broadened with width of $\eta=0.2$. 
For comparison, 
the coherent part in $\sigma(\omega)$ is also presented 
for the two cases of $N_{\rm s}=18$ 
by the broken lines of the same colors. 
}
\label{fig10}
\end{figure}

Finally, let us make comparisons with 
experimental results of optical conductivities. 
Recall here that the coherent and incoherent parts are not 
definitely separated 
in the experimentally observed optical conductivity. 
Thus, it is not so easy to make direct comparison 
between experimental and numerical $\sigma^{\rm reg}(\omega)$. 
Theoretically, on the other hand, 
one has to introduce the broadening of delta functions 
with some widths in dealing with finite systems 
as the present ones. 
Keeping these circumstances in mind, 
we consider the frequency dependence of $\sigma(\omega)$ 
including the Drude part;  
Figure \ref{fig10} depicts results of $U=10$, 
typical value of the Coulomb interaction for most 
of the high-$T_{\rm c}$ materials. 
For $\delta=0$, 
the present spectrum reveals a clear increase 
as $\omega\rightarrow 0$, 
even when the system is in the insulating state.  
This comes from the tiny $D$ 
due to the finite-size effect. 
For $\delta>0$, on the other hand, 
the Drude weight is considerably large in comparison with 
this tiny $D$ at half filling;  
there occurs a huge peak at $\omega=0$ 
as the coherent component. 
Its tail in Fig. \ref{fig10} appears 
due to the broadening of width $\eta=0.2$. 
Note here that even the tail is sizable 
in comparison with the intrinsic responses 
of $\sigma^{\rm reg}(\omega)$ in the inner-gap region. 
This means that in order to understand 
optical responses of the system, 
it is important to capture 
both the behavior of the coherent component 
and that of the incoherent one included 
in the total $\sigma(\omega)$ simultaneously. 
In Fig. \ref{fig10}, qualitatively the same change by doping 
in the transitions to the upper-Hubbard band is seen 
as that mentioned above for $U=20$ and 35. 
The structure in this $\omega$ region 
should be associated with the charge transfer 
from the copper $d$ to the oxygen $p$ states 
in the case of the cuprates, 
hence it is, in narrow sense, 
different from the present case of the Mott-Hubbard type. 
Aside from the character of the gap, i.e., 
either it is the Mott-Hubbard gap or the charge transfer gap, 
the evolution of doping observed 
in the experiments are correctly captured 
in our numerical results, 
the spectral weight transfer from the higher to lower energies 
as well as the collapse of the sharp peak 
at the low-energy edge. 

The evolution of the low-energy structure with doping 
contains both the growth of the Drude weight and 
the development of the incoherent mid-IR structure.  
The total sum of these two contributions 
is the effective carrier density $N_{\rm eff}$, 
whose weight is proportional to the doping concentration. 
In the doping process of the evolution from the insulator 
to low doping, 
the mid-IR incoherent part below the Mott gap grows quickly 
in our calculated results and is already prominent even 
in underdoped systems 
(see the result at $\delta\sim 0.111$), 
with a rather insensitive dependence on the further doping 
to the overdoped systems (see $\delta=0.333$).
On the contrary to the mid-IR part, the Drude part grows 
slowly at low doping, 
while it starts growing quickly from moderate to over doping, 
which is the origin of the scaling given by $D\propto \delta^2$ 
rather than $D\propto \delta$. 
The total sum resulting in $N_{\rm eff}$ evolves smoothly and 
linearly with $\delta$ 
because of the compensation of these two contributions.  
The fact that the Drude part remains only a small fraction 
of $N_{\rm eff}$ whereas the mid-IR structure 
rapidly grows from zero to low dopings is indeed common 
to all the high-$T_{\rm c}$ cuprates. 
It is even more universal in the transition metal oxides 
in general. 
Our calculated result for the further doping process 
from moderate to over doping is also consistent 
with the universal trend of the experimental results 
in transition metal oxides 
including the high-$T_{\rm c}$ cuprates, 
where the mid-IR structure is absorbed to the Drude weight. 
The $\delta^2$ scaling of the Drude weight is tightly related 
to this profound and universal trend 
in the doping process of the whole spectral weight 
in the optical conductivity.

\section{Conclusion and Discussion}

We have investigated the Drude weight 
of the Hubbard model on the square lattice by means 
of large-scale parallelized exact diagonalizations. 
The hole-doping dependence of the Drude weight 
near the transition point 
between the metal and the Mott insulator has been a 
controversial problem. 
We have very carefully examined finite-size effects 
and we have successfully captured the intrinsic behavior 
of the Drude weight. 
Our results suggest that 
the Drude weight is scaled by $D\propto \delta^{2}$,  
which is consistent with results from various kinds of analysis. 
In particular, it is important that 
this scaling agrees with the quantum Monte Carlo results 
through the scaling theory of the metal-insulator transition. 
The behavior $D\propto \delta^{2}$ 
is a characteristic feature 
near the marginal quantum critical point of the transition 
between the Mott insulator and the doped metal 
in two-dimensional systems. 
The restructuring of the electronic structure upon doping 
into the Mott insulator first causes 
the spectral weight transfer from the weight 
above the gap mainly to a mid-IR incoherent component 
below the gap. 
The spectral weight is further progressively transferred 
from the mid-IR component to the Drude weight. 
This large-scale restructuring seen in the frequency and 
doping dependences of $\sigma(\omega)$ in our result is 
consistent with the experimental results 
on transition metal oxides including high-$T_{\rm c}$ cuprates. 
The $\delta^2$ scaling of the Drude weight comes 
from this underlying physics involving the high energy scale. 

At $U=0$, the Hubbard model we studied has a perfect nesting 
just at half filling and 
the Fermi level coincides with the van Hove singularity. 
The effect of deviation from this perfect nesting 
by introducing the hopping to further sites 
(such as the next-nearest-neighbor hopping $t^{\prime}$) 
is left for separate study. 
The QMC study\cite{Furukawa_Imada_QMC_1993} indicates that 
the scaling behavior of the compressibility 
does not sensitively depend on the presence or absence 
of $t^{\prime}$, which implies that the present scaling 
of the Drude weight also holds. 
Next, let us recall the marginal point with $z=4$ 
in the one-dimensional Hubbard model 
with the next-nearest-neighbor 
hopping\cite{Nakano_Takahashi_MI_2004,
Nakano_Takahashi_Imada_CS_2005} mentioned above. 
In this case, a singularity in the density of states is absent 
at half filling at $U=0$ 
because the next-nearest-neighbor hopping is small. 
This fact also suggests that 
the metal-insulator transition with $z=4$ occurs 
irrespective of such a singularity in the density of states. 

The finite-size effects in such a quantity 
as the Drude weight are considerably large. 
Thus, the careful examination of the finite-size effects is 
necessary in possible ways as much as we can. 
If one calculates the quantity without considering the effects, 
obtained results could severely suffer 
from superficial finite-size effects. 
By such results, unfortunately, 
the incorrect conclusion would be derived.  
In this meaning, the present problem of the Drude weight 
is an instructive example. 
As a way to reduce the finite-size effect, 
the present work employs appropriately chosen clusters 
under the appropriate boundary conditions. 
Thereby, we have captured 
the critical behavior of $D\propto\delta^{2}$ 
near the half-filling insulator. 
Averaging over twisted boundary conditions\cite{Poilblanc} 
may be another possibility for the reduction. 
Quite recently 
such an investigation has been done 
in the $N_{\rm s}=20$ cluster 
of the $t$-$J$ model plus three-site terms\cite{Tohyama_JPCS}. 
Although this work gives 
a result of the $\delta$-dependence of $D$,  
it is so subtle that one cannot conclude 
either of $D\propto\delta^2$ or $D\propto\delta$. 
In the not-too-distant future, computers will 
be even more powerful; 
numerical diagonalizations of the Hubbard model 
with larger system sizes would be possible. 
Such calculations could clarify the validity 
of the present procedure for reducing the finite-size effects 
in $D$ and provide us with a more definite result 
for the critical behavior of $D$. 
We hope that 
the knowledge in the present study will serve 
for such future studies. 

\section*{Acknowledgements}
This work was supported by a Grant-in-Aid 
for Young Scientists (B) from the Ministry of Education, 
Culture, Sports, Science and Technology of Japan 
(15740221). 
A part of the computations was performed using facilities 
of the Information Initiative Center, Hokkaido University 
and the Supercomputer Center, 
Institute for Solid State Physics (ISSP), University of Tokyo. 
The authors thank Prof. Hajime Takayama and 
the staffs of the Supercomputer Center, ISSP 
for their support to perform our large-scale parallelized 
calculations in a special job class of the supercomputer of ISSP.


\begin{thebibliography}{9}

\bibitem{IFT_review} 
M. Imada, A. Fujimori and Y. Tokura: 
Rev. Mod. Phys. {\bf 70} (1998) 1039 and references therein. 
\bibitem{Dagotto_etal}  
\label{Dagotto_etal}  
E. Dagotto, A. Moreo, F. Ortolani, D. Poilblanc and J. Riera: 
Phys. Rev. B {\bf 45} (1992) 10741.
\bibitem{Nakano_Imada_1999}  
\label{Nakano_Imada_1999}  
H. Nakano and M. Imada: 
J. Phys. Soc. Jpn. {\bf 68} (1999) 1458.
\bibitem{Tohyama_etal}  
\label{Tohyama_etal}  
T. Tohyama, Y. Inoue, K. Tsutsui and S. Maekawa: 
Phys. Rev. B {\bf 72} (2005) 045113.
\bibitem{dimension_Hilbert_space} 
The largest dimension of the subspace of the Hubbard model 
with a given $N_{\rm s}$ is realized 
at half filling with the same number of 
up-spin electrons and down-spin ones. 
For examples, the dimensions are 
165~636~900 for $N_{\rm s}=16$, 
2~363~904~400 for $N_{\rm s}=18$ and   
34~134~779~536 for $N_{\rm s}=20$. 
It is possible to reduce these dimensions by using 
the symmetries of the Hamiltonian; however, 
the program for the reduced case is not necessarily applicable 
to any Hamiltonians having {\it arbitrary} shapes of clusters 
in contrast to our program. 
The versatile feature of our code makes it possible
to examine the various cases very easily. 
Although the dimension is very huge, 
our code costs 29.7 hours for the 260-time iterations 
to obtain the ground-state energy 
of the 20-site half-filling Hamiltonian for $U=35$, 
when we use 16 nodes 
in the supercomputer SR11000 in Hokkaido University. 
\bibitem{Cont_Frac_Method}  
E. Gagliano and C. Balseiro: 
Phys. Rev. Lett. {\bf 59} (1987) 2999.
\bibitem{Furukawa_Imada_QMC_1992}  
\label{Furukawa_Imada_QMC_1992}  
N. Furukawa and M. Imada: 
J. Phys. Soc. Jpn. {\bf 60} (1991) 3604; {\it ibid.}{\bf 61} (1992) 3331.
\bibitem{Nakano_Takahashi_PF_2003}  
H. Nakano and Y. Takahashi: 
J. Phys. Soc. Jpn. {\bf 72} (2003) 1191.
\bibitem{Nakano_Takahashi_SP_2004}  
H. Nakano and Y. Takahashi: 
J. Mag. Mag. Mat. {\bf 272-276} (2004) 487.
\bibitem{Nakano_Takahashi_MI_2004}  
H. Nakano and Y. Takahashi: 
J. Phys. Soc. Jpn. {\bf 73} (2004) 983.
\bibitem{comment_negative_D}
The negative Drude weights with large absolute values were 
reported at $\delta=0$ and $\delta=1/8$ for $N_{\rm s}=16$ 
in ref. \ref{Dagotto_etal}. 
In ref. \ref{Tohyama_etal}, 
the discontinuous dependence was reported 
at $\delta=1/3$ for $N_{\rm s}=18$. 
These works use only the periodic boundary condition. 
Note that such behavior occurs only at the hole concentration 
when the electronic shell structure is open. 
\bibitem{Tsunetsugu_Imada_tJ}  
H. Tsunetsugu and M. Imada: 
J. Phys. Soc. Jpn. {\bf 67} (1998) 1864.
\bibitem{Imada_scaling_1995}  
M. Imada: 
J. Phys. Soc. Jpn. {\bf 64} (1995) 2954.
\bibitem{Furukawa_Imada_QMC_1993}  
\label{Furukawa_Imada_QMC_1993}  
N. Furukawa and M. Imada: 
J. Phys. Soc. Jpn. {\bf 62} (1993) 2557.
\bibitem{Assaad_Imada_QMC_1996}  
F. Assaad and M. Imada: 
Phys. Rev. Lett. {\bf 76} (1996) 3176.
\bibitem{Imada_scaling_2005a}  
M. Imada: 
J. Phys. Soc. Jpn. {\bf 74} (2005) 859.
\bibitem{Imada_scaling_2005b}  
M. Imada: 
Phys. Rev. B {\bf 72} (2005) 075113.
\bibitem{Misawa_2006}  
T. Misawa, Y. Yamaji and M. Imada: 
J. Phys. Soc. Jpn. {\bf 75} (2006) 083705.
\bibitem{Nakano_Takahashi_Imada_CS_2005}  
H. Nakano, Y. Takahashi and M. Imada: 
Physica B {\bf 359-361C} (2005) 657.
\bibitem{LaSrCuO}  
S. Uchida, T. Ido, H. Takagi, T. Arima, Y. Tokura and S. Tajima: 
Phys. Rev. B {\bf 43} (1991) 7942.
\bibitem{Poilblanc}
D. Poilblanc: 
Phys. Rev. B {\bf 44} (1990) 9562.
\bibitem{Tohyama_JPCS}
T. Tohyama: J. Phys. Chem. Solids {\bf 67} (2006) 2210.
\end{thebibliography}
\end{document}